\documentclass[twocolumn,showpacs,superscriptaddress,pra]{revtex4}
\usepackage{amsmath}
\usepackage{graphicx}
\usepackage{amssymb}

\begin{document}

\title{Generalized Rashba spin-orbit coupling for cold atoms}

\date{\today{}}

\author{Gediminas Juzeli\=unas}

\affiliation{Institute of Theoretical Physics and Astronomy, Vilnius
University, A. Go\v{s}tauto 12, Vilnius LT-01108, Lithuania}

\author{Julius Ruseckas}

\affiliation{Institute of Theoretical Physics and Astronomy, Vilnius
University, A. Go\v{s}tauto 12, Vilnius F-01108, Lithuania}

\author{Jean Dalibard}

\affiliation{Laboratoire Kastler Brossel, CNRS, UPMC, Ecole Normale Sup\'erieure, 
24 rue Lhomond, 75005 Paris, France}
\begin{abstract}
We study the possibility for generating a new type of spin-orbit coupling for the
center-of-mass motion of cold atoms, using laser beams that resonantly couple
$N$ atomic internal ground states to an extra state. After a general analysis of
the scheme, we concentrate on the tetrapod setup ($N=4$) where the atomic state
can be described by a three-component spinor, evolving under the action of a
Rashba-Dresselhaus-type spin-orbit coupling for a spin 1 particle. We illustrate a
consequence of this coupling by studying the negative refraction of atoms at a
potential step, and show that the amplitude of the refracted beam is
significantly increased in comparison to the known case of spin $1/2$
Rashba-Dresselhaus
coupling. Finally we explore a possible implementation of this tetrapod setup,
using stimulated Raman couplings between Zeeman sublevels of the ground state of
alkali-metal atoms.
\end{abstract}


\maketitle

\section{Introduction}

The electron's spin degree of freedom plays a key role in the emerging area of
semiconductor spintronics \cite{Fert:2008,Grunberg:2008,Zutic04RMP}. A first
scheme for a semiconductor device is the spin field-effect Datta-Das transistor
(DDT). It was proposed 20 years ago \cite{dattadas} and implemented recently
\cite{Koo:2009Science}. Atomic and polaritonic analogs of the electron spin
transistor have also been suggested
\cite{Vaisnav08PRL-DDT,Johne09arxiv-Polariton-DDT}. An important ingredient of
the DDT is the spin-orbit coupling of the Rashba
\cite{Rashba60,Winkler03Review,Rashba-review08} or Dresselhaus
\cite{Dresselhaus55PR,Schliemann03PRL-DDT-Balanced} types. This
Rashba-Dresselhaus (RD) coupling scheme is described by a vector potential which can
be made proportional to the spin-$1/2$ operator of a particle within a plane
\cite{Schlieman06PRB}. It applies to electrons
\cite{Zutic04RMP,Winkler03Review,Rashba-review08} or atoms with two relevant
internal states
\cite{Dudarev04PRL,Ruseckas05PRL,Stanescu07PRL-Rashba,Jacob07APB,Juzeliunas08PRAR,Merkl09EPL-ZB,Vaishnav08PRL-ZB,Larson09PRA,Oh09-tripod}.

In the case of atoms, the spin-orbit coupling can be generated using two
counterpropagating light beams
\cite{Jacob07APB,Juzeliunas08PRAR,Merkl09EPL-ZB,Oh09-tripod} (or two standing
waves \cite{Stanescu07PRL-Rashba,Vaishnav08PRL-ZB,Larson09PRA}) and a third beam
propagating in an orthogonal direction, the beams being coupled to the atoms in
a tripod scheme \cite{Ruseckas05PRL,Unanyan98OC,Unanyan99pra}. The tripod atoms
have two degenerate internal dressed states known as \emph{dark states}, which
are immune to atom-light coupling. The center-of-mass motion of the dark-state
atoms is described by a two-component spinor and is equivalent to the motion of
a spin-$1/2$ particle with spin-orbit coupling
\cite{Stanescu07PRL-Rashba,Jacob07APB,Juzeliunas08PRAR,Merkl09EPL-ZB,Vaishnav08PRL-ZB,Larson09PRA,Oh09-tripod}
of the RD type.

\begin{figure}
\centering
\includegraphics[width=0.5\columnwidth]{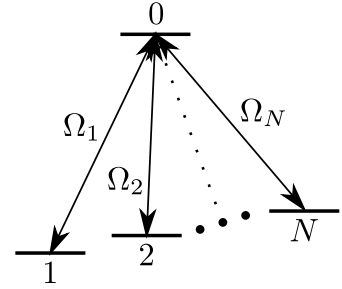}
\caption{$N$-pod configuration. An atomic state $|0\rangle$ is coupled to $N$
different atomic states $|j\rangle$, $j=1,...,N$ by $N$ resonant laser fields.}
\label{fig:figure-Npod}
\end{figure}

In the present article we investigate the possibility to generalize the
RD spin-orbit coupling scheme to spins larger than $1/2$. We
show that this can be achieved using cold atoms with more than two internal dark
states. We start our analysis with the general scheme in which $N$ laser beams
couple $N$ atomic internal ground states to a common excited state, thus forming
the $N$-pod setup shown in Fig.~\ref{fig:figure-Npod}. In the
$(N+1)$-dimensional Hilbert space, we identify $N-1$ dark states,
that is, zero-energy eigenstates of the atom-light Hamiltonian that are
superpositions of
the $N$ ground states and are immune to atom-light coupling. 

Subsequently, we analyze the tetrapod case ($N=4$) for which the center-of-mass
motion of the dark-state atoms is described by a three-component spinor and thus
corresponds to the motion of a spin-$1$ particle. We show that the resulting
spin-orbit coupling can be made of the RD type and yields three
cylindrically symmetric dispersion branches. Two of them are similar to those
for the familiar RD spin-$1/2$ Hamiltonian, so the atom can exhibit the
well-known
quasirelativistic behavior \cite{Juzeliunas08PRAR,Vaishnav08PRL-ZB} for
small wave vectors. Furthermore there is an extra branch with a flat dispersion
around zero momentum. The formation of the latter branch leads to interesting
phenomena, such as a possibility to have a negative refraction at a potential
step, characterized by a larger amplitude as compared to the spin-$1/2$ case.

Finally we explore a possible implementation of the tetrapod scheme for
alkali-metal
atoms using Raman transitions. To avoid a strong heating due to spontaneous
emission, all the states forming the tetrapod scheme are chosen among the Zeeman
sublevels of the atomic ground state, and are coupled by far-detuned Raman
lasers beams.

\section{The $N$-pod scheme}

\subsection{Atomic Hamiltonian}

We are interested in the center-of-mass motion of atoms in the field of several
light beams. The atoms are characterized by $N$ internal ground states
$|1\rangle$, $|2\rangle$, $\ldots$ , $|N\rangle$, which are resonantly coupled
to an extra state $|0\rangle$ by laser beams. This provides the $N$-pod
configuration shown in Fig.~\ref{fig:figure-Npod}. Note that the state
$|0\rangle$ does not necessarily represent an electronic excited level; it can
be a sublevel of the atomic ground state coupled to the states $|1\rangle$,
$|2\rangle$, $\ldots$ ,$|N\rangle$ via stimulated Raman transitions. A more
detailed discussion on practical implementation is presented in the
Sec.~\ref{sec:Production-of-the}.

The Hamiltonian describing the motion of an atom in the presence of the light
beams is
\begin{equation}
H_0 =\frac{\mathbf{p}^2}{2m}+V_{0}+V_1
\label{}
\end{equation}
where $m$ is the atomic mass and $\mathbf{p}=-i\hbar \nabla$ the atomic
momentum operator. The terms $V_0$ and $V_1$ describe the atom-light
interaction in the $N$-pod configuration and a possible additional external
potential, respectively. We assume for simplicity that all couplings
$|j\rangle \leftrightarrow |0\rangle$, $j=1,\ldots,N$ are resonant, so that
$V_0$ reads using the interaction representation and the rotating wave
approximation: 
\begin{equation}
V_0=\hbar\sum_{j=1}^N\Omega_j(\mathbf{r})\, |0\rangle\langle
j|+\mathrm{H.c.}\,,
\label{eq:H-0}
\end{equation}
where $\Omega_j$ is the Rabi frequency that couples the internal state
$|j\rangle$ to the common state $|0\rangle$, with $j=1,2,\ldots,N$. The
coupling $V_0$ can be rewritten as
\begin{equation}
V_0=\hbar\Omega(\mathbf{r})\Bigl[|0\rangle\langle
B(\mathbf{r})|+|B(\mathbf{r})\rangle\langle0|\Bigr]\,,
\label{eq:H-0-alternative}
\end{equation}
with 
\begin{equation}
|B\rangle=\frac{1}{\Omega}\sum_{
j=1}^N\Omega_j^*|j\rangle\,,\qquad\Omega^2=\sum_{j=1}^N|\Omega_j|^2\,.
\label{eq:B}
\end{equation}
Here $|B\rangle$ is the so-called bright (coupled) state and $\Omega$ is the
total Rabi frequency.

The  diagonalization of the atom-light interaction potential $V_0$ is straightforward:

(a) The coupling between the bright state $|B\rangle$ and the state $|0\rangle$ with
a strength equal to the Rabi frequency $\Omega$ in
Eq.~(\ref{eq:H-0-alternative}) gives rise to the two eigenstates
\begin{equation}
|\pm\rangle=\left(|B\rangle\pm|0\rangle\right)/\sqrt{2}\, ,
\label{eq:H-0-diagonal}
\end{equation}
with energies $\pm\hbar\Omega$.

(b) The remaining orthogonal $(N-1)$-dimensional subspace corresponds to dark states.
We denote $|D_n\rangle$, $n=1,\ldots, N-1$ an orthonormal basis of this
subspace. All dark states are eigenstates of the Hamiltonian $\hat{H}_0$ with
zero eigenenergy: $\hat{H}_0|D_n\rangle=0$ . They are orthogonal to the bright
state and to the state $|0\rangle$: $\langle B|D_n\rangle=\langle
0|D_n\rangle=0$.

Although the eigenenergies of the dark states are position-independent, the
states $|D_n\rangle$ depend on the atomic position through the spatial variation
of the Rabi frequencies $\Omega_j$. This leads to the appearance of the gauge
potentials to be considered next.

\subsection{Adiabatic motion of dark-state atoms}

We now suppose that the atoms are prepared in the dark-state subspace, and that
they move sufficiently slowly to remain in this manifold. This adiabatic
approximation is justified if the light fields are strong enough, so that the
energy difference $\pm \hbar\Omega$ between the dark-state manifold and the
other eigenstates $|\pm\rangle$ of $V_0$ is large compared to the detuning
due to Doppler shifts. The atomic state-vector $|\Phi\rangle$ can then be
expanded on the dark-state basis 
\begin{equation}
|\Phi\rangle=\sum_{j=1}^{N-1}\Psi_j(\mathbf{r})|D_j(\mathbf{r})\rangle,
\end{equation}
where $\Psi_j(\mathbf{r})$ is the wave function for the center-of-mass motion of
the atom in the $j$th dark state. The atomic center-of-mass motion is described
by an $(N-1)$-component wave function
\begin{equation}
\Psi=\left(
\begin{array}{c}
\Psi_1\\\ldots\\\Psi_{N-1}\end{array}\right)
\label{eq:psi-D-original}
\end{equation}
obeying the Schr\"odinger equation
\begin{equation}
i\hbar\frac{\partial}{\partial t}\Psi=H\Psi,
\label{eq:SE-reduced}
\end{equation}
with the Hamiltonian
\begin{equation}
H=\frac{1}{2m}(-i\hbar\nabla-\mathbf{A})^2+\Phi+V\,.
\label{eq:H}
\end{equation}
The potentials governing the atomic center-of-mass motion $\mathbf{A}$, $\Phi$,
and $V$ are $(N-1)\times(N-1)$ matrices. Here $\mathbf{A}$ and $\Phi$ are the
geometric potentials that emerge due to the spatial dependence of the atomic
dark states
\cite{Ruseckas05PRL,Berry84PRSA,wilczek84PRL,mead91,Bohm03Book,Shapere1989}.
The matrix $\mathbf{A}(\mathbf{r})$ represents a non-Abelian vector potential,
with the matrix elements
\begin{equation}
\mathbf{A}_{n,m}=i\hbar\langle D_n(\mathbf{r})|\nabla
D_m(\mathbf{r})\rangle\,,\quad n,m=1,\ldots,N-1\,.
\label{eq:A-nm}
\end{equation}
The matrix $\Phi(\mathbf{r})$ is an effective scalar potential known as the
Born-Huang potential. It can be expressed through the matrix elements of the
vector potential between the dark states and the bright state
$|B\rangle\equiv|D_0\rangle$:
\begin{equation}
\Phi_{n,m}=\frac{1}{2m}\mathbf{A}_{n,0}\cdot\mathbf{A}_{0,m}\,,\qquad
n,m=1,\ldots,N-1\,.
\label{eq:Phi-nm}
\end{equation}
The matrix $V(\mathbf{r})$ represents the restriction of $V_1(\mathbf{r})$ to
the dark state subspace. For simplicity we assume in the following that (i) the
matrix elements of $V_1$ between the dark-state manifold and the states
$|B\rangle$ or $|0\rangle$ are negligible, so that $V_1$ cannot cause any
significant departure of atoms from the dark-state manifold; (ii) $V$ is
proportional to the identity matrix in the dark state subspace, so that it does
not break the gauge symmetry of $(\mathbf{A},\Phi)$. For the particular case of
alkali-metal
atoms, this occurs when the trapping is provided by far-detuned laser beams. The
confinement potential is then the same for all sublevels of the electronic
ground state, in particular for the states $|j \rangle$ ($j=1,\,\ldots,\,N$)
considered here.

The non-Abelian vector potential $\mathbf{A}$ provides a curvature
(or effective ``magnetic'' field) 
\begin{equation}
\mathbf{B}=\nabla\times\mathbf{A}+\frac{1}{i\hbar}\mathbf{A}\times\mathbf{A}\ .
\label{eq:B-eff-initial}
\end{equation}
The first term represents the usual curl. Note that the second term
$\mathbf{A}\times\mathbf{A}$ does not vanish in general, since the Cartesian
components of the vector potential $\mathbf{A}$ do not necessarily commute
(i.e.\ the vector potential is non-Abelian). Therefore in contrast to the
Abelian case, even a constant vector potential can produce a nonzero curvature
and thus provide nontrivial topological effects, leading, for example, to unusual
dispersion curves.

\section{Effective fields generated by plane-wave laser beams }

\subsection{Dark states and gauge potentials}

From now on we focus on the case where the laser beams represent plane running
waves characterized by wave vectors $\mathbf{k}_j$, $j=1,\ldots,N$.
We suppose that the $N$
Rabi frequencies have equal amplitudes and read
\begin{equation}
\Omega_j=\frac{1}{\sqrt{N}}\Omega e^{i\mathbf{k}_j\cdot\mathbf{r}}\,,\qquad
j=1,2,\ldots,N\, .
\label{eq:Omega_j-plane-wave}
\end{equation}
At this stage the directions of the wave vectors $\mathbf{k}_j$ are still
arbitrary; we will address some specific geometries in Secs.~\ref{sec:planar}
and \ref{sec:tetrahedron}.

A convenient orthogonal set of $N-1$ normalized dark states is
\begin{equation}
|D_n\rangle=\frac{1}{\sqrt{N}}\sum_{j=1}^N|j\rangle e^{i2\pi
jn/N-i\mathbf{k}_j\cdot\mathbf{r}}\,,
\label{eq:D-n}
\end{equation}
with $n=1,2,\ldots,N-1$. Note that the bright state given by Eqs.~(\ref{eq:B})
and (\ref{eq:Omega_j-plane-wave}) has the form of Eq.~(\ref{eq:D-n}) with $n=0$,
so we will use in the following the notation $|D_0\rangle\equiv|B\rangle$.

Equations (\ref{eq:A-nm}) and (\ref{eq:D-n}) provide the matrix elements of the
vector potential
\begin{equation}
\mathbf{A}_{n,m}=\frac{\hbar}{N}\sum_{j=1}^N\mathbf{k}_je^{i2\pi
j(m-n)/N}\;.
\label{eq:A-nm-symmetric}
\end{equation}
It is evident that the vector potential $\mathbf{A}_{n,m}$ depends only on the
difference $n-m$, i.e.\ $\mathbf{A}_{n,m}=\mathbf{A}_{n-m,0}$.

Since the vector potential given by Eq.~(\ref{eq:A-nm-symmetric}) is constant in
space, the effective magnetic field (\ref{eq:B-eff-initial}) simplifies to
$i\hbar\mathbf{B}=\mathbf{A}\times\mathbf{A}$. Using
Eq.~(\ref{eq:A-nm-symmetric}), it can expressed in terms of the off-diagonal
matrix elements of the vector potential $\mathbf{A}_{n,0}$ and
$\mathbf{A}_{0,m}$ :
\begin{equation}
\mathbf{B}_{n,m}=\frac{i}{\hbar}\mathbf{A}_{n,0}\times\mathbf{A}_{0,m}\,.
\label{eq:B-eff-specific-nm}
\end{equation}

\subsection{Vector potential and angular momentum
\label{sub:Vector-potential-and}}

We now address the following question: Can the vector potential $\mathbf{A}$
be made proportional to a three-dimensional (3D) angular momentum operator
$\mathbf{J}$, that is,
$\mathbf{A}=\gamma\mathbf{J}$ , where $\gamma$ is a constant? If the answer
was positive, this would allow one to achieve a three-dimensional RD-type
coupling. This would be formally similar to the effective spin-orbit
interaction discussed in  \cite{Zygelman:1990}, arising from non-Abelian gauge
fields in molecular physics. However as we see now, one cannot use the present
scheme to achieve $\mathbf{A}\propto \mathbf{J}$.

The angular momentum operator is known to obey the following relations:
\begin{equation}
\mathbf{J}\times\mathbf{J}=i\hbar\mathbf{J}\, .
\label{eq:J-J-cross-product-J-relationship}
\end{equation}
If $\mathbf{A}=\gamma\mathbf{J}\,,$ the cross product of the vector potential
should be proportional to the vector potential itself:
$\gamma\mathbf{A}\times\mathbf{A}=i\hbar\mathbf{A}$ or simply
$\gamma\mathbf{B}=\mathbf{A}$. Using Eq.~(\ref{eq:B-eff-specific-nm}), the last
relationship would lead to 
\begin{equation}
\gamma\mathbf{A}_{n,0}\times\mathbf{A}_{m,0}^*=-i\hbar\mathbf{A}_{n
-m,0}\,,\quad n,m=1,\,\ldots,\,N-1\,.
\label{eq:C-cross-product}
\end{equation}
Multiplying Eq.~(\ref{eq:C-cross-product}) by $\mathbf{A}_{m,0}^*$, the
left-hand side of the resultant equation is zero. Thus one arrives at 
\begin{equation}
\mathbf{A}_{m,0}^*\cdot\mathbf{A}_{n-m,0}=0.
\label{eq:C-d-relation}
\end{equation}
Equation (\ref{eq:C-d-relation}) should hold for all possible values of $n$ and
$m$. In particular, by taking $m=1$ and $n=2$, one finds
$\mathbf{A}_{1,0}^*\cdot\mathbf{A}_{1,0}=0$. This equation can be fulfilled only
if $\mathbf{A}_{1,0}=0$. Then by taking $m=1$, the relationship
(\ref{eq:C-cross-product}) yields that
$\mathbf{A}_{n-1,0}=\mathbf{A}_{n+p,p+1}=0$ for integer $n$ and $p$. This means
that the vector potential $\mathbf{A}=\gamma\mathbf{J}$ should be identically
equal to zero.

In this way, we have proved that when using the $N$-pod scheme with plane waves of
equal amplitudes it is not possible to generate a nonzero vector potential
which is proportional to the 3D angular momentum operator $\mathbf{J}$. In other
words, it is not possible to produce a 3D spin-orbit coupling of the RD type
using the $N$-pod scheme. Yet one can get a two-dimensional (2D) RD coupling
by means of the
$N$-pod scheme. This includes not only the usual spin-$1/2$ RD coupling but
also a generalized 2D RD coupling for the spin-$1$ case, as we shall see
later on.

\section{Plane matter-wave solutions}

We suppose in the following that the external potential $V$ is uniform in space.
In this case the Schr\"odinger equation (\ref{eq:SE-reduced}) has plane-wave
solutions: 
\begin{equation}
\Phi_{\mathbf{k}}(\mathbf{r},t)=\Psi_{\mathbf{k}}e^{i\mathbf{k}\cdot\mathbf{r}
-\omega_{\mathbf{k}}t},
\label{eq:Phi-k-definition}
\end{equation}
where $\omega_{\mathbf{k}}$ is an eigenfrequency and $\Psi_{\mathbf{k}}$ is a
\textbf{$\mathbf{k}$}-dependent spinor: 
\begin{equation}
\Psi_{\mathbf{k}}=\left(
\begin{array}{c}
\Psi_{1,\mathbf{k}}\\\ldots\\\Psi_{(N-1),\mathbf{k}}\end{array}\right)\ .
\end{equation}
Note that the direction of the wave vector \textbf{$\mathbf{k}$} is arbitrary
and it is not related to the wave vectors of the light beams
\textbf{$\mathbf{k}_j$}.

The \textbf{$\mathbf{k}$}-dependent spinor $\Psi_{\mathbf{k}}$ obeys the
stationary Schr\"odinger equation 
\[
H_{\mathbf{k}}\Psi_{\mathbf{k}}=\hbar\omega_{\mathbf{k}}\Psi_{\mathbf{k}},
\]
with the $\mathbf{k}$-dependent Hamiltonian
\begin{equation}
H_{\mathbf{k}}=\frac{\hbar^2}{2m}k^2-\frac{\hbar}{m}\mathbf{A}\cdot\mathbf{k}
+\frac{1}{2m}\mathbf{A}^2+\Phi+V \ .
\label{eq:H-k}
\end{equation}
Exploiting Eqs.~(\ref{eq:Phi-nm}) and (\ref{eq:A-nm-symmetric}), the scalar term
$\mathbf{A}^2/2m+\Phi$ takes the form
\begin{equation}
\left(\frac{1}{2m}\mathbf{A}^2+\Phi\right)_{n,m}=\frac{\hbar^2}{2m}\frac{1}{
N}\sum_{j=1}^N\mathbf{k}_j^2e^{i\frac{2\pi}{N}(m-n)j}\,.
\label{eq:A-square
+ Phi-1}
\end{equation}
If the wave vectors of all the Rabi frequencies have the same modulus
$\mathbf{k}_j^2=2\kappa^2$, the term 
\begin{equation}
\frac{1}{2m}\mathbf{A}^2+\Phi=\frac{\hbar^2\kappa^2}{m}\hat{I}
\end{equation}
is proportional to the unit matrix $\hat{I}$ for any arrangement of the
wave vectors (both planar and 3D). In this case the Hamiltonian (\ref{eq:H-k})
simplifies to
\begin{equation}
H_{\mathbf{k}}=\frac{\hbar}{2m}
\Bigl(\hbar k^2-2\mathbf{A}\cdot\mathbf{k}+2\hbar \kappa^2\Bigr)+V\,.
\end{equation}
If the external trapping potential $V$ is proportional to the unit matrix, the
eigenvectors $\Psi_{\mathbf{k}}^{\beta}$ of the Hamiltonian $H_{\mathbf{k}}$ are
also the eigenvectors of the operator
$A_{\mathbf{k}}=\mathbf{A}\cdot\mathbf{k}/k$ representing the projection of the
vector potential along the wave vector,
\begin{equation}
A_{\mathbf{k}}\Psi_{\mathbf{k}}^{\beta}=-\hbar\kappa\beta\Psi_{\mathbf{k}}^{
\beta}\,,
\label{eq:A-k-eigenstate}
\end{equation}
where the dimensionless parameter $\beta\equiv\beta_{\mathbf{k}}$ depends
generally on the wave-vector $\mathbf{k}$. The corresponding eigenvalues of the
Hamiltonian $H_{\mathbf{k}}$ are
\begin{equation}
\hbar\omega_{\mathbf{k}}^{\beta}=\frac{\hbar^2}{2m}\Bigl[(k+\beta\kappa)^2+(2
-\beta^2)\kappa^2\Bigr]+V.
\label{eq:omega-k-beta}
\end{equation}
For $\mathbf{k}=\mathbf{0}$ all the eigenenergies
$\hbar\omega_{\mathbf{k}}^{\beta}$ are
equal and do not depend on the branch parameter $\beta$. Consequently all
dispersion branches merge to $\omega_{0}^{\beta} \equiv \omega_0$ at the origin
where $k=0$. To find the eigenstates and the eigenenergies for $k\ne0$, one
needs to specify the arrangement of the wave vectors $\mathbf{k}_j$.

\section{Planar geometry}
\label{sec:planar}

\subsection{Wave vectors on a regular polygon}

Let us analyze a situation where the wave vectors $\mathbf{k}_j$ are situated in
a plane and form a regular polygon
\begin{eqnarray}
\mathbf{k}_j & = &
\sqrt{2}\kappa[-\cos\alpha_j\mathbf{e}_x+\sin\alpha_j\mathbf{e}_y]\\ &
= & -\kappa\Bigl(e^{i\alpha_j}\mathbf{e}_{+}+e^{-i\alpha_j}\mathbf{e}_{-}\Bigr)\,,
\end{eqnarray}
with $\mathbf{e}_{\pm}=\frac{1}{\sqrt{2}}(\mathbf{e}_x\pm i\mathbf{e}_y)$, where
$\alpha_j=2\pi j/N$ is the angle between the wave vector and the $x$ axis. The
scalar and vector potentials, Eqs.~(\ref{eq:Phi-nm}) and
(\ref{eq:A-nm-symmetric}), take the form 
\begin{eqnarray}
\Phi_{n,m} & = & \frac{\hbar^2\kappa^2}{2m}(\delta_{m,1}\delta_{n,1}+\delta_{m,N
-1}\delta_{n,N-1})\,,
\label{eq:Phi-general-result} \\
\mathbf{A}_{n,m} & = & -\hbar\kappa\sum_{\pm}\mathbf{e}_{\pm}\delta_{n,m\pm1}\,.
\label{eq:A-nm-poligon}
\end{eqnarray}
The vector potential is thus a tridiagonal matrix whose elements are
proportional to $\mathbf{e}_x\pm i\mathbf{e}_y$, whereas the scalar potential
$\Phi_{n,m}$ is a diagonal matrix with nonzero elements only for $n=m=1$ or
$n=m=N-1$.

Note that the matrices $A_x$ and $A_y$, are proportional to the $x$ and $y$
components of the angular momentum operator $\mathbf{J}$ only for the tripod
($N=3$) and tetrapod ($N=4$) schemes. In these cases the scalar potential is
proportional to $J_z^2$.

The projection of $\mathbf{A}_{n,m}$ along the wave vector is
\begin{equation}
(A_{\mathbf{k}})_{n,m}=-\frac{\hbar\kappa}{\sqrt{2}}\Bigl(\delta_{n,m+1}e^{i\varphi}
+\delta_{n,m-1}e^{-i\varphi}\Bigr)\,,
\end{equation}
where $\varphi$ is the angle between the wave vector $\mathbf{k}$ and the $x$
axis. The eigenvectors of this operator are
\begin{equation}
\Psi_{\mathbf{k}}^{\beta}=\sqrt{\frac{2}{N}}\left(
\begin{array}{c}
\sin\left(\frac{\pi q}{N}\right)\\\sin\left(2\frac{\pi
q}{N}\right)e^{i\varphi}\\\cdots\\\sin\left((N-1)\frac{\pi
q}{N}\right)e^{i(N-2)\varphi}\end{array}\right)\,,
\label{eq:Phi-k-beta-N}
\end{equation}
with $q=1,\dots N-1$. The corresponding eigenvalues are given by
Eq.~(\ref{eq:A-k-eigenstate}) with
\begin{equation}
\beta=\sqrt{2}\cos\left(\frac{\pi q}{N}\right)\,.
\label{eq:beta}
\end{equation}
It is to be emphasized that the dimensionless parameter $\beta$ does not depend
on $\mathbf{k}$ for this particular geometry. The vectors
$\Psi_{\mathbf{k}}^{\beta}$ represent eigenstates of the Hamiltonian with
eigenenergies $\omega_k^{\beta}$ given by Eqs.~(\ref{eq:omega-k-beta}) and
(\ref{eq:beta}). This provides $N-1$ dispersion branches.

\subsection{Tripod setup}

\begin{figure}
\centering
\includegraphics[width=0.9\columnwidth]{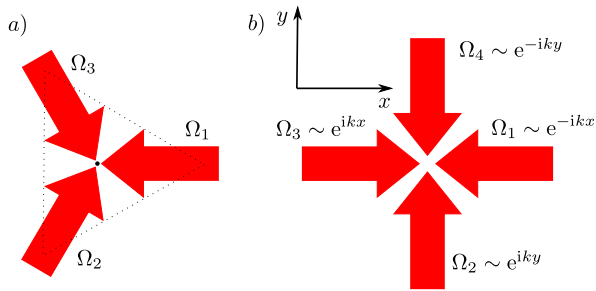}
\caption{(Color online) Planar arrangement of laser beams for tripod (a) and
tetrapod (b) setups.}
\label{fig:tripod-directions}
\end{figure}
Consider first the tripod setup ($N=3$) in which the wave vectors $\mathbf{k}_j$
form an equilateral triangle [Fig.~\ref{fig:tripod-directions}(a)]. The parameter
$\beta$ featured in Eqs.~(\ref{eq:A-k-eigenstate}) and (\ref{eq:beta}) then
takes the values $\hbar\beta/\sqrt{2}=\pm\hbar/2$, representing the eigenvalues
of the projection of a spin-$1/2$ on a given axis. In such a situation the
operator $\mathbf{A}$ is related to the spin $1/2$ operator
$\hbar\boldsymbol{\sigma}_{\bot}$, providing the RD coupling along the $xy$
plane as in the previous studies
\cite{Stanescu07PRL-Rashba,Jacob07APB,Juzeliunas08PRAR,Merkl09EPL-ZB,Vaishnav08PRL-ZB,Larson09PRA,Oh09-tripod}:
\begin{equation}
\mathbf{A}=-\hbar\kappa\boldsymbol{\sigma}_{\bot}/\sqrt{2}\; .
\label{eq:A-tripod}
\end{equation}
It is noteworthy that the present setup produces a cylindrically symmetric
spin-orbit coupling in a more straightforward manner than the previously
suggested tripod schemes. Those schemes involved two counterpropagating light
beams \cite{Jacob07APB,Juzeliunas08PRAR,Merkl09EPL-ZB,Oh09-tripod} (or two
standing waves \cite{Stanescu07PRL-Rashba,Vaishnav08PRL-ZB,Larson09PRA}) and a
third beam propagating in an orthogonal direction. Consequently, one needed to
add a detuning potential and make the amplitudes of the Rabi frequencies
asymmetric in order to have dispersion curves of the RD-type, with the
proper cylindrical symmetry
\cite{Stanescu07PRL-Rashba,Jacob07APB,Juzeliunas08PRAR,Merkl09EPL-ZB,Vaishnav08PRL-ZB,Larson09PRA,Oh09-tripod}.
On the contrary, for the present regular polygon arrangement of wave vectors,
the dispersion relation is naturally symmetric as long as the amplitudes of all
four Rabi frequencies are equal.

\subsection{Tetrapod setup\label{sub:Tetrapod-seup}}

For $N=4$ one arrives at the tetrapod setup involving two pairs of
counterpropagating laser fields shown in Fig.~\ref{fig:tripod-directions}(b). In
this case the vector potential reads 
\begin{equation}
\mathbf{A}=\frac{\hbar\kappa}{\sqrt{2}}\left(
\begin{array}{ccc}
0 & -\mathbf{e}_x+i\mathbf{e}_y & 0\\ -\mathbf{e}_x-i\mathbf{e}_y & 0 &
-\mathbf{e}_x+i\mathbf{e}_y\\ 0 & -\mathbf{e}_x-i\mathbf{e}_y & 0
\end{array}\right)\,.
\label{eq:A-tetrapod-matrix}
\end{equation}
The possible values for the parameter $\beta$ featured in
Eq.~(\ref{eq:A-k-eigenstate}) are $\hbar\beta=0,\,\pm\hbar$, representing the
eigenvalues of the component of a spin $1$ along a given axis. Consequently the
operator $\mathbf{A}$ is proportional to the projection $\mathbf{J}_{\bot}$ of a
spin $1$ operator along the $xy$ plane
\begin{equation}
\mathbf{A}=-\kappa\mathbf{\mathbf{J}_{\bot}}\,,\qquad\mathbf{J}_{
\bot}=J_x\mathbf{e}_x+J_y\mathbf{e}_y\, .
\label{eq:A-tetrapod}
\end{equation}
The scalar potential can be represented in terms of the $z$ component of the spin operator
\begin{equation}
\Phi=\frac{\hbar^2\kappa^2}{2m}J_z^2\,.
\label{eq:Phi-tetrapod}
\end{equation}

\begin{figure}
\centering
\includegraphics[width=0.9\columnwidth]{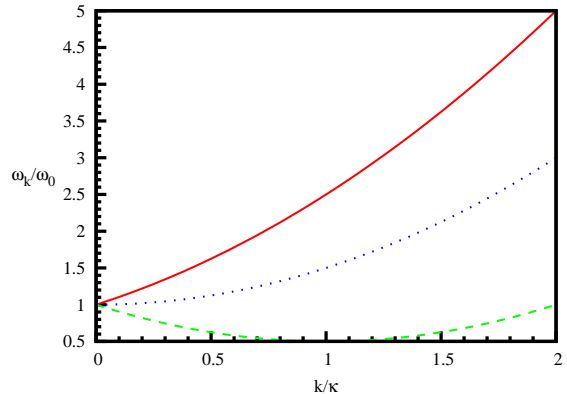}
\caption{(Color online) Dispersion curves for the tetrapod scheme calculated
using Eq.~(\ref{eq:omega-k-beta-quatro-pod}) for $V=0$. Here
$\omega_0=\hbar \kappa^2/m$.}
\label{fig:dispersion}
\end{figure}

The eigenstates and the eigenenergies of the Hamiltonian are now
\begin{equation}
\Psi_{\mathbf{k}}^{\pm1}=\frac{1}{2}\left(
\begin{array}{c}
1\\\pm\sqrt{2}e^{i\alpha}\\ e^{2i\alpha}
\end{array}\right)\,,\quad\Psi_{\mathbf{k}}^0=\frac{1}{\sqrt{2}}\left(
\begin{array}{c}
-1\\ 0\\ e^{2i\alpha}\end{array}\right)\,,
\label{eq:Psi-k-tetrapod}
\end{equation}
and
\begin{equation}
\hbar\omega_k^{\beta}=\frac{\hbar^2}{2m}\Bigl(k^2+2\kappa
k\beta+2\kappa^2\Bigr)+V\,,\quad\beta=0\,,\pm1.
\label{eq:omega-k-beta-quatro-pod}
\end{equation}
For $\beta=\pm1$ the dispersion curves shown in Fig.~\ref{fig:dispersion} are
analogous to those of the spin-$1/2$ RD model. An additional dispersion curve
with $\beta=0$ represents a parabola centered at $k=0$. 

The dispersion curve with $\beta=-1$ has its minimum at
$\hbar\omega=\hbar^2\kappa^2/2m$, whereas the other two dispersion branches have
minima at the double energy $\hbar\omega=\hbar^2\kappa^2/m$ (for $V=0$).
Therefore, all dispersion curves have a strictly positive minimum energy. This
nonzero minimum originates from the micromotion of the atom in the light
field, caused by nonadiabatic transitions between the dark and bright states
\cite{Aharonov:1992,Cheneau08EPL}. The associated kinetic energy gives rise to
the scalar potential given by Eq.~(\ref{eq:Phi-general-result}), which has a
nonzero contribution even when acting on the dark states. 

Finally, we note an important difference in the ``topology'' of the
eigenfunctions for the RD spin-$1/2$ and spin-$1$ problems, even thought the
$\beta=\pm1$ branches have the same dispersion in the two cases: The
wave functions $\Psi_{\mathbf{k}}^{\beta}$ exhibit a $\pi$ Berry's phase in $k$
space in the spin-$1/2$ case, whereas this Berry's phase is absent for the
spin-$1$.

\section{Tetrahedron geometry}

\label{sec:tetrahedron}

In this section we present an example of a nonplanar setup, which has some
advantages with respect to the planar configuration investigated in the previous
section, because it leads to a simpler scalar potential. We consider again the
tetrapod setup ($N=4$) with wave vectors $\mathbf{k}_j$ arranged in a regular
tetrahedron geometry:
\begin{equation}
\hat{\mathbf{k}}_j\cdot\hat{\mathbf{k}}_{j^{\prime}}=-\frac{1}{3}\,,\quad
j\ne j^{\prime}\,.
\label{eq:tetrahedron-condition}
\end{equation}
where $\hat{\mathbf{k}}_j=\mathbf{k}_j/k_j$ is a unit vector. More precisely,
we choose
\begin{equation}
\mathbf{k}_{1,3}  = \kappa^{\prime}(\pm \mathbf{e}_y\sqrt{2}-\mathbf{e}_z)\;,\quad
\mathbf{k}_{2,4}  = \kappa^{\prime}(\pm\mathbf{e}_x\sqrt{2}+\mathbf{e}_z)\;.
\label{}
\end{equation}
Using Eq.~(\ref{eq:A-nm-symmetric}), the vector potential then reads
\begin{equation}
\mathbf{A}=\frac{\hbar\kappa^{\prime}}{\sqrt{2}}\left(
\begin{array}{ccc}
0 & -\mathbf{e}_x+i\mathbf{e}_y &\sqrt{2}\mathbf{e}_z\\
-\mathbf{e}_x-i\mathbf{e}_y & 0 & -\mathbf{e}_x+i\mathbf{e}_y\\
\sqrt{2}\mathbf{e}_z & -\mathbf{e}_x-i\mathbf{e}_y & 0\end{array}\right)\;.
\label{eq:A-tetrahedron}
\end{equation}
For atoms moving in the $xy$ plane the vector potential can be expressed in
terms of a spin-$1$ operator in the $xy$ plane:
$\mathbf{A}_{\bot}=-\kappa^{\prime}\mathbf{J}_{\bot}$. Hence we obtain as before
a RD-type spin-orbit coupling for the atomic motion in the $xy$ plane,
characterized by the dispersion relation shown in Fig.~\ref{fig:dispersion}. Yet
we are now dealing with a 3D problem, so the same dispersion also characterizes
the atomic motion along two other planes perpendicular to the vectors
$\mathbf{e}_x+\mathbf{e}_y$ and $\mathbf{e}_x-\mathbf{e}_y$. By making an atomic
lattice along these directions, the atomic tunneling will be influenced by a
spin-$1$ RD coupling, thus extending the previous studies of spin-$1/2$
RD coupling in lattices \cite{Goldman09PRA}. This will be investigated in a
separate study.

A distinguished feature of the tetrahedron geometry is that the scalar potential
is proportional to the unit matrix $\hat{I}$:
\begin{equation}
\Phi=\frac{\hbar^2\kappa^{\prime2}}{2m}\hat{I}\;.
\label{eq:Phi-tetrahedron}
\end{equation}
Thus for atoms placed in a 3D lattice, there is no energy mismatch between
different dark states located in adjacent sites. This contrasts with the planar
tetrapod case (Eq.~(\ref{eq:Phi-tetrapod})), where the spin components are
likely to get frozen in the lattice because tunneling matrix elements are
normally much smaller than the atomic recoil energy, which gives the scale for
the scalar potential.  

It is noteworthy that the $z$ component of the vector potential given by
Eq.~(\ref{eq:A-tetrahedron}) is not proportional to $J_z$. Hence, one cannot
generate a 3D Hamiltonian with RD-type spin-orbit coupling for all
directions of the atomic motion. This is a particular case of the general
conclusion reached in the Sec.~\ref{sub:Vector-potential-and}. 

\section{Transmission by a potential step}

A spectacular consequence of spin-orbit RD coupling is the negative
refraction and reflection that occurs when a matter wave is incident on a
potential step. The problem was investigated for spin-$1/2$ atoms
\cite{Juzeliunas08PRAR,juz08-neg-refl} and electrons \cite{Winkler09PRB}. In
this case one can calculate relatively easily the transmission and reflection of
the atomic wave packet. For small wave vectors of the incident atoms,
$k\ll\kappa$, the transmission probability is close to unity at zero angle of
incidence. Here the parameter $\kappa$ characterizes the strength of the
spin-orbit interaction, see Eq.~(\ref{eq:A-tripod}). This nearly complete
transmission is a manifestation of the Klein paradox appearing also for electron
tunneling in graphene \cite{Katsnelson06NP}. For a nonzero angle of incidence,
the transmission probability is less than $1$ and decreases with increasing
angle. Furthermore the transmitted matter wave experiences negative refraction
\cite{Juzeliunas08PRAR}, similar to the case of electrons in graphene
\cite{Cheianov07Science}.

Particles with a spin larger than $1/2$ have additional degrees of freedom,
which modifies the continuity conditions at the potential step. This can lead to
a significant increase of the transmission probability of atoms, as we show now
for particles submitted to a spin-$1$ RD coupling. 

\subsection{The Hamiltonian}

We consider in this section the motion of a particle in the $xy$ plane described
by the Hamiltonian
\begin{equation}
H=\frac{1}{2m}\left(\hat{\mathbf{p}}^2+2\hbar\kappa\hat{\mathbf{p}}\cdot\mathbf{
J}_{\bot}+2\hbar^2\kappa^2\right)+V(x)
\label{eq:H-Spin-1}
\end{equation}
where $\mathbf{J}_{\bot}=J_x\mathbf{e}_x+J_y\mathbf{e}_y$ is the projection of
spin-$1$ operator onto the $xy$ plane. Such a Hamiltonian can be obtained using
the tetrapod setups described in the Secs.~\ref{sub:Tetrapod-seup} and
\ref{sec:tetrahedron}. The external potential $V(x)$ is given by the step
function along $x$
\begin{equation}
V(x)=
\begin{cases}
0, & x\leq 0\\ V_0, & x>0\end{cases}
\label{eq:V(x)}
\end{equation}
with $V_0>0$. It is convenient to introduce the wave vector
$k_0=2mV_0/\hbar^2\kappa$ characterizing the height of the barrier.

For a constant potential the eigenvalue equation has plane-wave solutions
(\ref{eq:Phi-k-definition}) characterized by the spinor part
$\Psi_{\mathbf{k}}^{\beta}$ [Eq.~(\ref{eq:Psi-k-tetrapod})]. The corresponding
eigenvalues $\hbar\omega_k^{\beta}$ are given by
Eq.~(\ref{eq:omega-k-beta-quatro-pod}) with $k=\sqrt{k_x^2+k_y^2}$ and are
plotted in Fig.~\ref{fig:dispersion}. Additionally there can be evanescent wave
solutions localized in the vicinity of the potential step in the $x>0$ region.
In that case we have $k_x=iq$ with $q>0$, giving 
\begin{equation}
k=\sqrt{k_y^2-q^2}\,,\qquad q^2<k_y^2\,.
\label{eq:k-evanesc}
\end{equation}
For the present problem, only the evanescent wave with $\beta=0$ will play a
role, 
\begin{equation}
\Psi_{k_y,q}^0=c_{k_y,q}^0\left(
\begin{array}{c}
-1\\ 0\\\frac{q+k_y}{q-k_y}\end{array}\right)\,,
\label{eq:Evanescent-beta--0}
\end{equation}
where $c_{k_y,q}^{\beta}$ is the normalization factor.

\subsection{Incident waves with $\beta=1$}

\begin{figure}
\centering
\includegraphics[width=0.7\columnwidth]{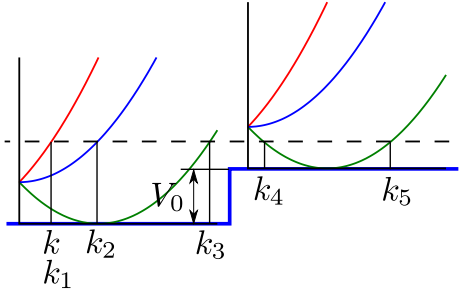}
\caption{(Color online) Wave numbers of reflected and transmitted waves and
energy conservation at the potential step.} 
\label{fig:dispersion-barrier}
\end{figure}

\begin{figure}
\centering
\includegraphics[width=0.7\columnwidth]{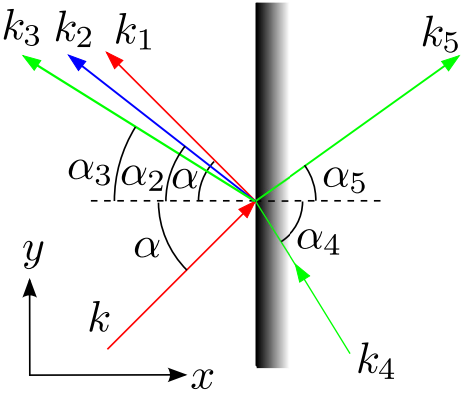}
\caption{(Color online) Reflection and transmission of atoms at a potential
step. In addition there is an evanescent transmitted wave with
$\mathbf{k}_6=k_y\mathbf{e}_y+iq\mathbf{e}_x$.} 
\label{fig:geomery}
\end{figure}

In this paragraph we restrict our analysis to the case where the incident atom
is prepared in the upper dispersion branch ($\beta=1$) in the region $x<0$.
Denoting its wave vector by $\mathbf{k}$, the incident wave is
\begin{equation}
\Psi_{\mathrm{in}}=\Psi_{\mathbf{k}}^1e^{i\mathbf{k}\cdot\mathbf{r}}\;.
\label{eq:Psi-in}
\end{equation}
The potential step is assumed to be high enough, 
\begin{equation}
k_0>k^2/\kappa+2k\,,
\label{eq:V-0--Condition}
\end{equation}
so that there can be no propagating transmitted waves with chirality $\beta=0$
or $\beta=1$ (see Fig.~\ref{fig:dispersion-barrier}). At the same time, to allow
for propagation of plane waves in the region $x>0$ for the lower dispersion
branch $\beta=-1$, the step height should not be too large:
\begin{equation}
k_0<\kappa+k^2/\kappa+2k\,.
\end{equation}
The directions of reflected and transmitted waves are depicted in
Fig.~\ref{fig:geomery}. The reflected waves generally contain all three
components,
\begin{equation}
\Psi_{\mathrm{refl}}=r_1\Psi_{\mathbf{k}_1}^1e^{i\mathbf{k}_1\cdot\mathbf{r}}
+r_2\Psi_{\mathbf{k}_2}^0e^{i\mathbf{k}_2\cdot\mathbf{r}}+r_3\Psi_{\mathbf{
k}_3}^{-1}e^{i\mathbf{k}_3\cdot\mathbf{r}}\,,
\label{eq:Psi-refl}
\end{equation}
where $k_1=k$, $k_2=\sqrt{k^2+2\kappa k}$ and $k_3=k+2\kappa$. The reflection
angles are $\pi-\alpha$, $\pi-\alpha_2$, and $\pi-\alpha_3$, with
$\alpha_2=\arcsin[\sin(\alpha)k/k_2]$ and $\alpha_3=\arcsin[\sin(\alpha)k/k_3]$.
The transmitted waves are
\begin{equation}
\Psi_{\mathrm{tr}}=t_4\Psi_{\mathbf{k}_4}^{-1}e^{i\mathbf{k}_4\cdot\mathbf{r}}
+t_5\Psi_{\mathbf{k}_5}^{-1}e^{i\mathbf{k}_5\cdot\mathbf{r}}+t_6\Psi_{\mathbf{
k}_6}^0e^{i\mathbf{k}_6\cdot\mathbf{r}}\,,
\label{eq:Psi-tr}
\end{equation}
where $k_4=\kappa-\sqrt{(k+\kappa)^2-k_0\kappa}$,
$k_5=\kappa+\sqrt{(k+\kappa)^2-k_0\kappa}$ , and $k_6=k^2+2\kappa k-k_0\kappa$
(see Fig.~\ref{fig:dispersion-barrier}). The first and second transmitted waves
experience negative and positive refraction, respectively, and propagate at the
angles $\pi-\alpha_4$ and $\alpha_5$, where
$\alpha_4=\arcsin[\sin(\alpha)k/k_4]$ and $\alpha_5=\arcsin[\sin(\alpha)k/k_5]$.
On the other hand, due to the condition (\ref{eq:V-0--Condition}) the third
transmitted wave with the helicity $\beta=0$ is an evanescent one along the $x$
axis and thus is characterized by the wave vector
$\mathbf{k}_6=k_y\mathbf{e}_y+iq\mathbf{e}_x$, with $k_y=k\sin\alpha$ and
$q=\sqrt{k_y^2-k_6^2}$. Note that there is no evanescent transmitted wave in the
upper dispersion branch ($\beta=1$) because it cannot comply with the momentum
conservation along the interface in addition to the energy conservation. 

The multicomponent wave function and its first derivative in the $x$ direction
are required to be continuous at the barrier ($x=0$), providing six equations
containing six unknown coefficients $r_1$, $r_2$, $r_3$, $t_4$, $t_5$, and $t_6$.
Of special interest is the situation where $k_0=4k$. In this case the
wave number of the first refracted wave coincides with the wave number of the
incident wave, $k_4=k$, so the angle of refraction is equal to the angle of
incidence for the first reflected wave. $\alpha_4=\alpha$.

The analytical solution for the six coefficients is generally complicated. It is
instructive to obtain approximate solutions for small wave vectors and small
angles of incidence, $k\ll\kappa$ and $\alpha\ll1$. In such a case one can
restrict to reflected (\ref{eq:Psi-refl}) and transmitted (\ref{eq:Psi-tr})
waves containing only the contributions of $\mathbf{k}_1$, $\mathbf{k}_2$,
$\mathbf{k}_4$, and $\mathbf{k}_6$. The transmitted wave with $\mathbf{k}_6$
represents a rapidly decaying evanescent wave characterized by a spinor
component given by Eq.~(\ref{eq:Evanescent-beta--0}) with $q\gg k_y$:
\begin{equation}
\Psi_{\mathbf{k}_6}^0\approx\frac{1}{\sqrt{2}}\left(
\begin{array}{c}
-1\\ 0\\ 1\end{array}\right)\;.
\label{eq:g-k6-0}
\end{equation}
The continuity of the wave function at $x=0$ gives
\begin{equation}
\Psi_{\mathbf{k}}^1+r_1\Psi_{\mathbf{k}_1}^1+r_2\Psi_{\mathbf{k}_2}^0e^{
i\mathbf{k}_2\cdot\mathbf{r}}=t_4\Psi_{\mathbf{k}_4}^{-1}+a\Psi_{\mathbf{k}_6}^0\;.
\label{eq:continuity-approx-2}
\end{equation}
In addition, we require continuity of the derivative in the $x$ direction for
the component with $\beta=0$, which is the largest:
\begin{equation}
k_2\cos(\pi-\alpha_2)r_2\Psi_{\mathbf{k}_2}^0=iqa\Psi_{\mathbf{k}_6}^0
\label{eq:continuity-derivatives-approx-2}
\end{equation}
Here $k_4\approx k_0/2-k$ and $\alpha_2\approx\sqrt{k/2\kappa}\sin(\alpha)$, with
$k_2=\sqrt{2\kappa k}$ and $q\approx\sqrt{\kappa(k_0-2k)}$. Using the spinors
(\ref{eq:Psi-k-tetrapod}) and (\ref{eq:g-k6-0}) one obtains the following
solution to Eqs.~(\ref{eq:continuity-approx-2}) and
(\ref{eq:continuity-derivatives-approx-2}):
\begin{eqnarray}
t_4 & = &
\frac{2\cos\alpha}{\cos\alpha+\cos\alpha_4}e^{i(\alpha+\alpha_4)}
\label{eq:t4-approx-2}
\\ r_1 & = &
\frac{\cos\alpha_4-\cos\alpha}{\cos\alpha+\cos\alpha_4}e^{i2\alpha}
\label{eq:r1-approx-2}
\\ r_2 & = &
\frac{\sqrt{k_0-2k}}{\sqrt{k}+i\sqrt{k_0/2-k}}\cos\alpha\tan\left(\frac{\alpha
+\alpha_4}{2}\right)e^{i\alpha}
\label{eq:r2-approx-2}
\end{eqnarray}
The calculated reflection and transmission coefficients $r_1$ and $t_4$ obey the
probability conservation up to terms of the order of $O(\alpha^2)$:
\begin{equation}
|r_1|^2+\frac{\cos\alpha_4}{\cos\alpha}|t_4|^2\approx1\,.
\label{eq:probability-conservation-1}
\end{equation}
If the barrier height is such that $k_0=4k$, we have $\alpha_4=\alpha$. In that
case $|t_4|\approx1$ and $|r_1|\approx0$ leading to an almost perfect negative
refraction, at the exactl opposite refraction angle, provided $k\ll\kappa$ and
the angles of incidence are not too large. 

\begin{figure}
\centering
\includegraphics[width=0.9\columnwidth]{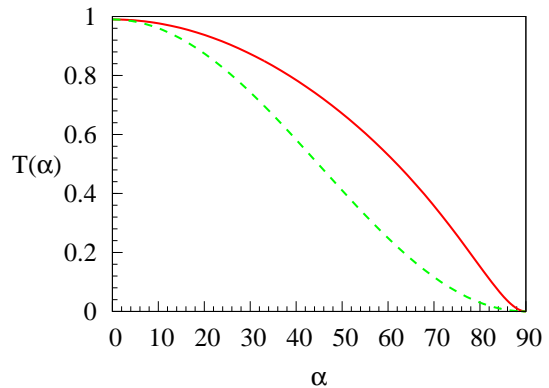}
\caption{(Color online) Transmission probability of negatively refracted atoms
at a potential step as a function of the angle of incidence $\alpha$ for a
spin-$1$ (solid red) and spin-$1/2$ (dashed green) systems. The parameters used
for the calculation are $k/\kappa=0.1$ and $k_0/\kappa=0.4$ for both systems.}
\label{fig-compar-1-0.5}
\end{figure}

Figure~\ref{fig-compar-1-0.5} presents the comparison of the transmission
probabilities for the spin-$1$ and spin-$1/2$ RD coupling using the exact
numerical solutions of the continuity equations at the boundary $x=0$. For the
spin-$1/2$ case the incident wave is also prepared in the upper dispersion
branch. The figure shows a marked increase in the transmission probability for
small angles of incidence in the case of spin $1$. Note that the transmitted
waves experience negative refraction both for the spin-$1$ and the
spin-$1/2$ cases.

\section{Implementation of the tetrapod setup with alkali-metal atoms}

\label{sec:Production-of-the}

\begin{figure}
\centering
\includegraphics[width=0.6\columnwidth]{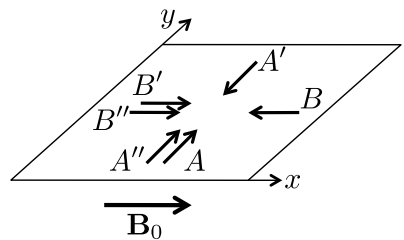}
\caption{Directions of the laser beams used for implementing the tetrapod
coupling scheme with alkali-metal atoms, via stimulated Raman transitions.}
\label{fig:exp-directions}
\end{figure}

\begin{figure}
\centering
\includegraphics[width=1\columnwidth]{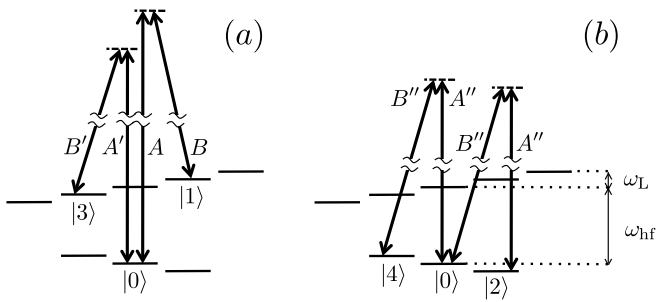}
\caption{Implementation of the tetrapod scheme for alkali-metal atoms with two
hyperfine levels of angular momentum $F=1$ and $F=2$. The laser couplings
involved in this scheme correspond to stimulated Raman transitions between
hyperfine  states of the ground atomic level. We choose
$|0\rangle\equiv|F=1,m_F=0\rangle$. (a) The laser beams $A$, $A^{\prime}$, $B$
and $B^{\prime}$ induce the transitions
$|0\rangle\to|1\rangle\equiv|F=2,m_F=1\rangle$ and
$|0\rangle\to|3\rangle\equiv|F=2,m_F=-1\rangle$. (b) The laser beams
$A^{\prime\prime}$ and $B^{\prime\prime}$ induce the transitions
$|0\rangle\to|2\rangle\equiv|F=1,m_F=1\rangle$ and
$|0\rangle\to|4\rangle\equiv|F=1,m_F=-1\rangle$. }
\label{fig:exp_scheme}
\end{figure}

We now discuss a possible implementation of the tetrapod scheme. We consider the
case of alkali-metal atoms, which are the most frequently used in current experiments.
In order to avoid a strong heating due to spontaneous emission, we study the
case where the state $|0\rangle$ is actually one of the Zeeman sublevels of the
ground state. The states $|j\rangle$ (with $j=1,\ldots,4$) are also Zeeman
sublevels of the ground state, and the coupling between the state $|0\rangle$
and a state $|j\rangle$ is provided by a pair of laser beams that induce a Raman
transition under the condition of the two-photon resonance. The use of Raman
transitions in this context is an extension to the tetrapod case of a recent
proposal \cite{Spielman09PRL} to implement a $\Lambda$-type scheme for the
generation of an effective magnetic field by means of the counterpropagating
laser beams \cite{Cheneau08EPL,Juzeliunas06PRA}.

We recall that the electronic ground level $nS_{1/2}$ of alkali-meal atoms is split
by hyperfine interaction in two sublevels with angular momenta $F=I+1/2$ and
$F=I-1/2$, where $I$ is the nuclear spin. We consider in the following the case
$I=3/2$ that is relevant for lithium ($^7\mathrm{Li}$, $n=2$), sodium
($^{23}\mathrm{Na}$, $n=3$) or rubidium ($^{87}\mathrm{Rb}$, $n=5$). In order to
minimize the rate of spontaneous emission processes, we restrict to Raman
transitions that are far detuned from the resonance with the ``true'' excited
states $nP_{1/2}$ or $nP_{3/2}$ of the $D_1$ or $D_2$ transitions. More
precisely the typical one-photon detuning of the beams involved in the Raman
process is chosen much larger than the hyperfine structure of the excited level
$nP_{1/2}$ or $nP_{3/2}$ ($0.8\,\mathrm{GHz}$ for the hyperfine splitting of the
level $5P_{1/2}$ of $^{87}\mathrm{Rb}$). At the same time the one-photon
detuning should be smaller than the fine structure splitting, that is, the
difference between the energies of $nP_{1/2}$ and $nP_{3/2}$
($7000\,\mathrm{GHz}$ for $^{87}\mathrm{Rb}$). When the one-photon detuning
exceeds the hyperfine splitting, the nucleus angular momentum does not play any
role in the selection rules that determine the allowed transitions for photon
absorption or emission. For the $D_1$ ($D_2$) transition, the allowed
couplings are the same as between a spin-$1/2$ ground level and a spin-$1/2$
($3/2$) excited level. In particular, the only allowed Raman transitions
correspond to a change $\Delta m_J=0$ or $\Delta m_J=\pm1$ of the azimuthal
quantum number $m_J$.

A scheme that fulfills the aforementioned constraints is represented in
Figs.~\ref{fig:exp-directions} and \ref{fig:exp_scheme}. The atomic motion along
the $z$ direction is supposed to be frozen thanks to a trapping potential
$m\omega_z^2z^2/2$ such that $\hbar\omega_z$ is much larger than the atomic
kinetic energy. The atom is placed in a uniform magnetic field $B_0$ directed
along the $x$ direction. The role of this magnetic field is to allow for a
selective Raman excitation between two given Zeeman sublevels. More precisely
the Larmor frequency $\omega_{\mathrm{L}}=\mu_{{\rm B}}B_0/\hbar$ ($\mu_{{\rm
B}}$ is the Bohr magneton) is chosen much larger than the two-photon (Raman)
Rabi frequency $\Omega$. Typically we choose $\omega_{\mathrm{L}}/2\pi$ on the
order of a few MHz (i.e.\ $B_0$ on the order of a few Gauss) and $\Omega/2\pi$
in the range $10^5-10^6\,\mathrm{Hz}$. The latter choice is sufficient to ensure
that the splitting $\hbar\Omega$ between the dark-state manifold and the states
$|\pm\rangle$ is large compared to the two-photon Doppler shift, as required for
the adiabatic approximation to be valid. The state $|0\rangle$ is chosen equal
to the $|F=1,m_F=0\rangle$ sublevel and the states $|j\rangle$ with
$j=1,\ldots,4$ are the $|F=2,m_F=\pm1\rangle$ and $|F=1,m_F=\pm1\rangle$
sublevels. Here the quantization axis is the $x$ axis, parallel to the direction
of the magnetic field $\mathbf{B}_0$. As indicated in
Fig.~\ref{fig:exp_scheme}(a), the transition between $|0\rangle$ and
$|1\rangle\equiv|F=2,m_F=+1\rangle$ is driven by a pair of laser beams $(A,B)$
with a frequency difference equal to
$\omega_{\mathrm{hf}}+\omega_{\mathrm{L}}/2$, where $\omega_{\mathrm{hf}}$ is
the hyperfine splitting between the $F=1$ and $F=2$ manifolds
($\omega_{\mathrm{hf}}/2\pi$ is on the order of $7\,\mathrm{GHz}$ for
$^{87}\mathrm{Rb}$). The laser beam $A$ propagates along the $y$ axis (wave
vector $k{\bf\mathbf{e}_y}$, where ${\bf\mathbf{e}_y}$ is a unit vector). It is
linearly polarized along $x$, so that it carries no angular momentum along the
$x$ axis. The laser beam $B$ propagates along the $x$ axis (wave vector
$-k{\bf\mathbf{e}_x}$) and is circularly ($\sigma_{-}$) polarized. In the
transition $|0\rangle\to|1\rangle$ the momentum change of the atom is $\hbar{\bf
k_1}=\hbar k({\bf\mathbf{e}_x}+{\bf\mathbf{e}_y})$. One can readily check that
the transition $|0\rangle\to|1\rangle$ is the only one that is driven resonantly
by this pair of beams, thanks to the fact that the Land\'e factors are opposite
for the $F=1$ and $F=2$ manifolds, as one can see in Fig.~\ref{fig:exp_scheme}(a).
Similarly, the transition between $|0\rangle$ and
$|3\rangle\equiv|F=2,m_F=-1\rangle$ is driven by a pair of laser beams $(A',B')$
with a frequency difference equal to
$\omega_{\mathrm{hf}}-\omega_{\mathrm{L}}/2$. The beam $A'$ propagates along $y$
with wave vector $-k{\bf\mathbf{e}_y}$ and is linearly polarized along $x$. The
beam $B'$ propagates along $x$ with wave vector $k{\bf\mathbf{e}_x}$ and is
circularly ($\sigma_{+}$) polarized. The atomic momentum change in the
transition $|0\rangle\to|3\rangle$ is $\hbar{\bf k_3}=-\hbar{\bf k_1}$. The
difference in the frequencies of $A$ and $A'$ is chosen large enough so that no
transition is driven with a significant probability by the pairs of beams
$(A,B')$ and $(A',B)$.

The two remaining states of the tetrapod configuration are
$|2\rangle\equiv|F=1,m_F=+1\rangle$ and $|4\rangle\equiv|F=1,m_F=-1\rangle$. The
coupling between these states and the state $|0\rangle$ is provided by a single
pair of laser beams $(A'',B'')$, as in the recent experiment
\cite{Spielman09PRL}, in which the $\Lambda$ (ladder) type coupling was
generated within the Zeeman sublevels of the $F=1$ manifold. The wave vector of
$A''$ is $k{\bf\mathbf{e}_y}$ and this beam is linearly polarized along $x$. The
beam $B''$ propagates along $x$ with wave vector $k{\bf\mathbf{e}_x}$ and is
circularly ($\sigma_{+}$) polarized. The frequency difference between the beam
$A''$ and $B''$ is $\omega_{\mathrm{L}}/2$ so that the pair $(A'',B'')$
resonantly drives the transition $|0\rangle\to|2\rangle$ with a momentum
transfer $\hbar{\bf k_2}=\hbar k({\bf\mathbf{e}_x}-{\bf\mathbf{e}_y})$, and the
transition $|0\rangle\to|4\rangle$ with a momentum transfer $\hbar{\bf
k_4}=-\hbar{\bf k_2}$. Note that here again we take advantage of the different
signs of the Land\'e factors of the $F=1$ and $F=2$ manifolds: The pair of beams
$(A'',B'')$ cannot resonantly drive a transition between two sublevels of the
$F=2$ manifold []see Fig.~\ref{fig:exp_scheme}(b)]. Note also that another
consequence of two-photon processes is a modification of the energies of the
states $|j\rangle$, via the absorption and stimulated emission of photons in the
same laser beam. It can be accounted for by including these energy shifts in the
choice of the two-photon detunings and, for example, by taking advantage of the
(small) second-order Zeeman shift.

This configuration therefore constitutes a suitable implementation of the scheme
discussed in the first part of this article. The momentum transfers $\hbar{\bf
k_j}=\hbar k(\pm{\bf\mathbf{e}_x}\pm{\bf\mathbf{e}_y})$ form a square in the
$xy$ plane shown in Fig.~\ref{fig:tripod-directions}(b) (subject to the rotation
of the coordinate system by $45^{\circ}$). The intensities of the various beams can be
adjusted so that all Rabi frequencies $\Omega_j$ are equal, once the
Clebsch-Gordan coefficients associated to each Raman transition have been taken
into account (note that the two-photon Rabi frequencies for
$|0\rangle\to|2\rangle$ and $|0\rangle\to|4\rangle$ transitions are equal by
construction). With a one-photon detuning of $3\,\mathrm{nm}$ , which represents
$1/5$ of the fine structure splitting for rubidium atoms, the residual photon
scattering rate is below $1\,\mathrm{s}^{-1}$ for a two-photon Rabi frequency
$\Omega/(2\pi)=10^5\,\mathrm{Hz}$. The corresponding heating rate is thus small
enough to provide enough time for the investigation of the RD coupling
studied in this article.

\section{Conclusions}

In this article we have explained how to produce a spin-orbit coupling of the
RD type for a spin larger than $1/2$. Our scheme makes use of
cold atoms with three or more internal dark states so that their quasi-spin is
equal to or greater than unity. We have analyzed a general scheme in which $N$
laser beams couple $N$ atomic internal ground states to an extra state, thus
forming an $N$-pod setup of light-matter interaction. In this case the atoms
have $N-1$ dark states representing superpositions of the $N$ ground states that
are immune to the atom-light coupling.

We have analyzed in detail the particular case of the tetrapod setup ($N=4$), in
which the center of mass motion of the atoms in their dark state manifold is
described by a three-component spinor and thus corresponds to the motion of a
spin-$1$ particle. We have shown that the resulting spin-orbit coupling can be
made of the RD type and yields three cylindrically symmetric dispersion
branches. Two of them are similar to those known for the familiar RD spin-$1/2$ Hamiltonian, so the atom can exhibit a quasirelativistic behavior
\cite{Juzeliunas08PRAR,Merkl09EPL-ZB,Vaishnav08PRL-ZB} for small wave vectors.
Furthermore, we have shown that there exists an extra branch with a flat
dispersion around zero momentum. We have studied the modifications that this
extra branch brings to the problem of negative refraction of matter waves on a
potential step, and shown that it enhances the negative refraction probability. 

Finally we have discussed a possible implementation of the tetrapod setup with
cold alkali-metal atoms. We have shown that in order to avoid heating due to
spontaneous emission, it is possible to choose all the states involved in this
tetrapod scheme among the various Zeeman sublevels of the ground atomic state.
All laser couplings are then provided by stimulated Raman transitions. For
rubidium atoms, realistic parameters yield a residual spontaneous emission rate
below $1\,\mathrm{s}^{-1}$, which makes the observation of this spin-orbit
coupling scheme experimentally feasible.

\begin{acknowledgments}
We thank M. Lewenstein and S. Das Sarma for helpful discussions. This work has
been supported by the Gilibert program, the Lithuanian Science and Studies
Foundation (Grant No.~V-34/2009), the Research Council of Lithuania, the R{\'e}gion
Ile de France IFRAF, the ANR (Grant No.~ANR-08-BLAN-65 BOFL), and the EU projects
SCALA and  STREP NAMEQUAM. LKB is a mixed research unit No.8552 of CNRS, ENS, and Universit{\'e}
Pierre et Marie Curie.
\end{acknowledgments}

\end{document}